\documentclass{emulateapj}
\usepackage{lscape}

\slugcomment{To appear in the Publications of the Astronomical Society of the Pacific}
\shorttitle{Photometry with IRAC}
\shortauthors{Hora et al.}

\begin{document}

\title{Photometry using the Infrared Array Camera on the
Spitzer Space Telescope}

\author{ Joseph L. Hora\altaffilmark{1}, Sean Carey\altaffilmark{2}, Jason
Surace\altaffilmark{2}, Massimo Marengo\altaffilmark{1}, Patrick
Lowrance\altaffilmark{2}, William J. Glaccum\altaffilmark{2}, Mark
Lacy\altaffilmark{2}, William T. Reach\altaffilmark{2}, William F.
Hoffmann\altaffilmark{3}, Pauline Barmby\altaffilmark{4}, S. P.
Willner\altaffilmark{1}, Giovanni G. Fazio\altaffilmark{1}, S. Thomas
Megeath\altaffilmark{5}, Lori E. Allen\altaffilmark{1}, Bidushi
Bhattacharya\altaffilmark{6}, Manuel Quijada\altaffilmark{7}}

\altaffiltext{1}{Harvard-Smithsonian Center for Astrophysics, 60 Garden Street, MS-65,
Cambridge, MA 02138}
\altaffiltext{2}{{\it Spitzer} Science Center, MS 220-6,
California Institute of Technology, Pasadena, CA 91125}
\altaffiltext{3}{Steward Observatory, Univ. of Arizona, Tucson, AZ 85721}
\altaffiltext{4}{Department of Physics \& Astronomy,
University of Western Ontario,
1151 Richmond St,
London, ON N6A 3K7, CANADA}
\altaffiltext{5}{Ritter Observatory, mail drop 113,
	The University of Toledo,
	2801 West Bancroft Street,
	Toledo, Ohio 43606}
\altaffiltext{6}{NASA/Herschel Science Center, California Institute of Technology, Pasadena, CA 91125}
\altaffiltext{7}{Goddard Space Flight Center, Optics Branch, Code 551,
Greenbelt, MD 20771}

\email{jhora@cfa.harvard.edu}

\begin{abstract}

We present several corrections for point source photometry to be applied
to data from the Infrared Array Camera (IRAC)
on the Spitzer Space Telescope.  These corrections are necessary because
of characteristics of the IRAC arrays and optics and the way the
instrument is calibrated in-flight.  When these corrections are applied,
it is possible to achieve a $\sim$2\% relative photometric accuracy for
sources of adequate signal to noise in an IRAC image.


\end{abstract}

\section{Introduction}

The Infrared Array Camera (IRAC) was built at the NASA Goddard Space
Flight Center under the direction of a team led by Giovanni Fazio at the
Smithsonian Astrophysical Observatory \citep{fazio04}. IRAC is the
mid-infrared camera on the Spitzer Space Telescope \citep{werner}, with
four arrays, or ``channels'', simultaneously taking data in two separate
fields of view. The four channels are referred to in this paper with their
standard labels of 3.6, 4.5, 5.8, and 8.0 $\mu$m for channels 1, 2, 3, 4,
respectively, although as described by the IRAC
documentation\footnote{http://ssc.spitzer.caltech.edu/irac/} and by
\cite{fazio04,reach05} the nominal wavelengths differ from these labels.  
The absolute calibration of the camera was performed in flight by
comparison to a set of stars that had been selected and characterized
before launch \citep{megeath03,cohenxiii}. \cite{reach05} presented the
IRAC in-flight calibration results, including the observing strategy, the
predictions and measurements, and an assessment of the calibration
accuracy and stability of the instrument and the pipeline-processed data
provided by the Spitzer Science Center (SSC) to observers.

There are several characteristics of the IRAC instrument that affect the
accuracy of the photometry obtained from the images.  The effects
considered in this paper are:

\begin{enumerate}
\item The IRAC science pipeline generates
images in units of surface brightness, but
because of distortion, the pixels do not subtend constant solid
angle.  This causes errors in point source
photometry that vary over the field of view (FOV) in each channel.
\item The IRAC spectral response varies over the field of view, and
therefore the color corrections are field dependent.
\item The electron rates in the 3.6 $\mu$m and 4.5 
$\mu$m channels depend slightly on
pixel phase (the position of the star relative to the nearest pixel
center).  The pipeline calibration factors are correct on average,
as is appropriate for sources observed multiple times at multiple
dither positions.  For the most precise photometry, however, pixel
phase should be taken into account.
\item A correction must be applied for the size of the aperture and 
background region used in aperture photometry, if different from that 
used by \citet{reach05} to derive the calibration from standard stars.
\end{enumerate}

The corrections described in this paper should be applied to the
photometry in a manner consistent with those applied by \citet{reach05}
(which includes using the centroiding technique and the aperture and
annulus background sizes described by Reach et al., applying the point
source gain correction described in Section 2.2 below, and the pixel phase
correction described in Section 3 below), in order for the absolute
calibration to remain valid and to achieve $<$2\% photometric accuracy
reported.

\section{Field-of-View (FOV) dependent effects}

\begin{figure}
\begin{center}
\includegraphics[scale=0.65,viewport=10 200 590 780,clip=true,angle=0]{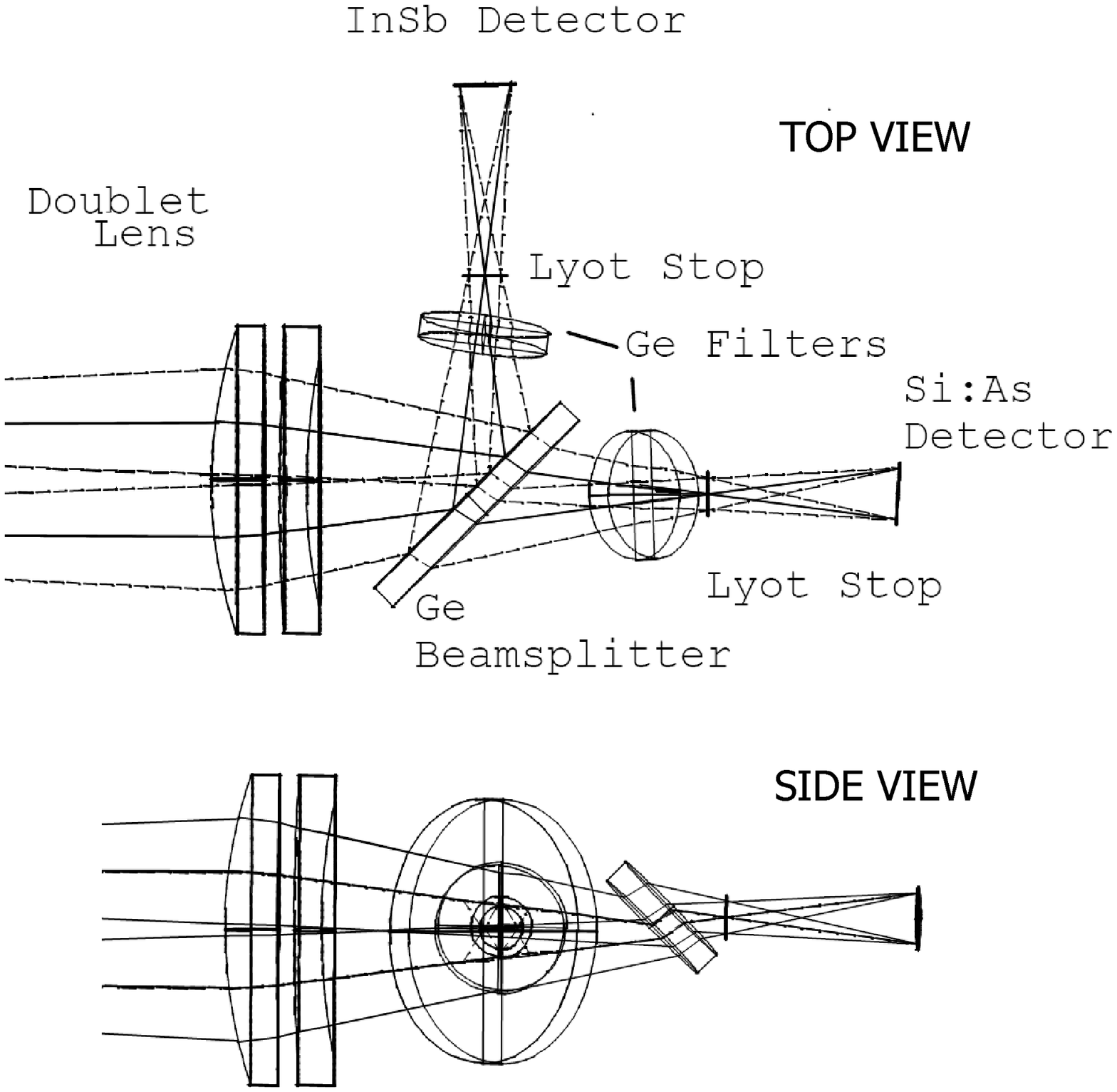}
\figcaption{
The IRAC optical design layout (interior to the camera body), showing the
side view and top view.  There are two fields viewed simultaneously, with
channels 1 and 3 viewing one field and channels 2 and 4 the other.  In
each pair, the light is reflected from the surface of the beamsplitter and
passes through a filter to the InSb detector (channels 1 and 2).  The
longer wavelength light passes through the beamsplitter and filters to the
Si:As detectors (channels 3 and 4). The range of angles of incidence on
the filters and beamsplitters depending on position in the field of view
is apparent from the rays traced through the system. \label{optfig}}
\end{center}
\end{figure}

A portion of the IRAC optical layout is shown in Figure \ref{optfig}, and
the details are described by \citet{fazio04}.  This figure shows the
relationship between the channels -- the two separate fields of view are
shared by pairs of channels (channels 1 \& 3 and channels 2 \& 4).  Each
pair shares the same doublet lens and is divided at the beamsplitter where
the short wavelength light is reflected and long wavelength transmitted.  
In both channels, the light then passes through filters before entering
the detector arrays at an angle that depends on the position in the field
of view. The tilted elements and the variable angle of incidence on the
filters and array causes some of the effects on the IRAC photometry
described below.

\clearpage

\subsection{Distortion and Pixel Area}

\begin{figure}
\epsscale{0.7}
\plotone{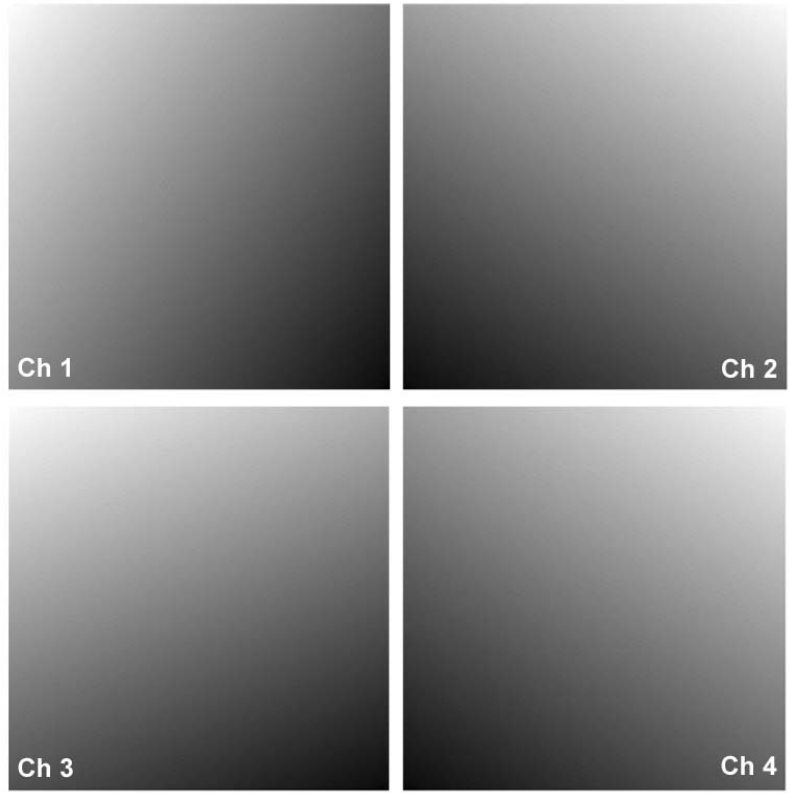}
\caption{
Distortion correction images for each of the IRAC channels.  Darker
regions represent regions where response to a point source has a lower
value than in the lighter regions. The full range of variations are 2.5\%,
3.2\%, 2.4\%, and 3.8\% for channels 1-4, respectively. The arrays are
shown in ``BCD orientation'', with the first pixel in the FITS file shown
in the lower left of each image, with the most rapidly varying index from
left to right. \label{distort}}

\end{figure}

The IRAC images have distortion in each channel due to its optical design,
resulting in a pixel displacement of $\sim$2 pixels or less in the corners
of the array relative to a regularly-spaced grid aligned with the central
pixel \citep{fazio04}.  The distortion also causes the pixel area to vary
slightly over the FOV.  The distortion was measured using data taken
during In-Orbit Checkout (IOC) period shortly after launch. The distortion
can be fit by a quadratic model, which is incorporated into the BCD image
headers \citep{shupe05}. The change in pixel area over the FOV was
estimated by calculating the determinant of the Jacobian of the
transformation to the distorted coordinate systems \citep{sparks00}.  The
results are shown in Figure \ref{distort}.  The total ranges of the pixel
area changes are 2.5\%, 3.2\%, 2.4\%, and 3.8\% for channels 1-4,
respectively.

\subsection{Point Source Gain Correction}

The pipeline gain or ``flat'' correction is determined from observations
of the zodiacal background (zodi) emission.  Regions of high and low zodi
background (near the ecliptic plane and poles, respectively)  are observed
during an IRAC campaign.  The flat image in each band is obtained by
rejecting point sources in the dithered frames and differencing the high
and low zodi images.  The illumination by the zodi is assumed to be
uniform over the FOV, so dividing each science image by a normalized
version of the flat corrects most of the pixel-to-pixel gain
variations that exist in the arrays.  The flat correction has been found
not to vary over time during the mission within the measurement noise, so
data from the whole Spitzer mission have been combined into a
``superflat'' which is used to correct the data for the entire mission.  
The SSC will soon reprocess all of the IRAC data with the S18 pipeline,
which will use the superflat constructed from the first four years of
Spitzer operation.

\begin{figure}
\begin{center}
\includegraphics[scale=0.45,angle=-90]{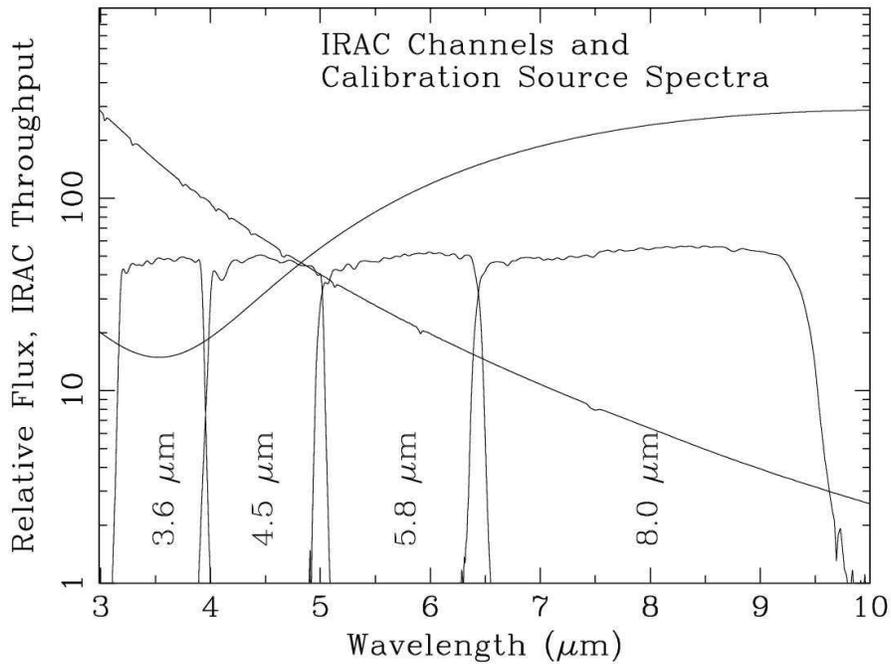}
\caption{The relative band transmissions of the four IRAC channels are
shown compared to the spectra of an A0V standard star, and a model of the
zodi emission in the ecliptic pole region \citep{kelsall98}. The vertical
axis is the logarithm of the total instrumental transmission
\citep{fazio04}, or for the star and zodi it is the logarithm of the flux
density (W cm$^{-2} \mu$m$^{-1}$) scaled to fit on the
plot.\label{calstar}}
\end{center}
\end{figure}

The flat correction works very well for extended sources with colors
similar to the zodi.  The changes in pixel area over the FOV due to the
geometric distortion are part of the flat, as are the effective
wavelength variations, and the BCD are calibrated to units of MJy/sr. In
addition, the spectrum of the zodi is quite different than that of a
typical star (Figure \ref{calstar}), so essentially the opposite
correction for effective wavelength is applied by the flat to data from
normal stars. Furthermore, scattered light from extended emission outside
of the FOV could affect the flat measurement near the array edges, and
light or charge spreading within the array could cause differences between
extended and point source photometry. All of these effects and possibly 
others lead to a variation in the photometry of a point source at 
different locations on the array.  We can use the photometry of a star at
many points on the array to derive a correction to apply to remove the 
variation.

\subsubsection{Data and Analysis}

We derived the point source gain correction from characterization data
taken during IOC.  The observations were taken after telescope cooldown
and final focus adjustment.  A standard star (BD+67 1044 $=$ SAO 17718)
was observed on a 5x5 grid (a square grid with equal spacings of roughly
50 pixels) across the arrays using Astronomical Observing Requests (AORs)  
\dataset{ADS/Sa.Spitzer\#0006946816} and
\dataset{ADS/Sa.Spitzer\#0006946560}.  This is a K2 star that had the
brightness in the IRAC bands (magnitudes of 6.29, 6.43, 6.39, and 6.33 in
channels 1-4, respectively) such that with relatively short frame times
(0.4 sec for channels 1 and 2, and 2 sec for channels 3 and 4) the peak
pixel could be kept near the middle of the linear range of the detectors
and a high S/N measurement of the focus across the FOV could be performed.
The short frame times also minimized the number of cosmic rays in the
image. At each of the positions of the 5x5 grid, a set of 12 small dithers
was performed to minimize the effects of bad pixels or any pixel gain map
problems.

\begin{figure}
\epsscale{0.7}
\plotone{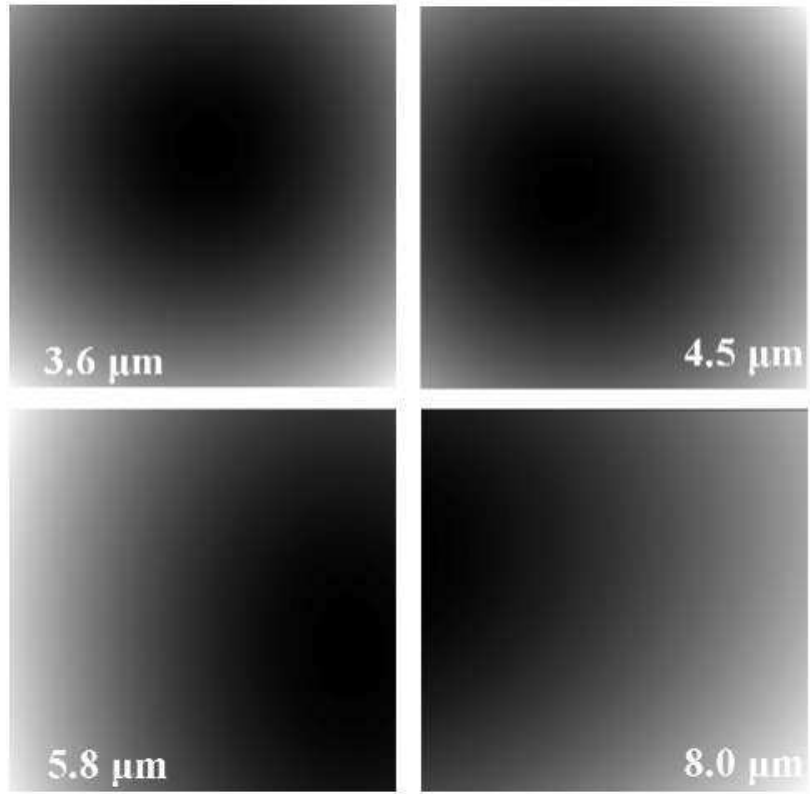}
\caption{
The point source photometry correction images for each IRAC band.  
Darker regions represent regions where the photometry of a point source 
has a lower value than in the lighter regions.  The photometry therefore
should be divided by the value given in the images to correct the measurement
relative to the center of the array.\label{pointcorr}
}
\end{figure}

From each image, the stellar flux was extracted using the {\tt phot}
command in IRAF\footnote{IRAF is distributed by the National Optical
Astronomy Observatory, which is operated by the Association of
Universities for Research in Astronomy (AURA) under cooperative agreement
with the National Science Foundation.}. A radius of 10 IRAC pixels was
used.  In each channel, the photometry varies systematically across the
FOV.  The variations can be fit by a quadratic surface across the arrays,
a different one for each IRAC channel.  The point source photometry
correction factor $F_{psp}$ was found by fitting the function
\begin{equation} 
F_{psp}(x,y) = A + B(x-128) + C(y-128) + $$
$$D(x-128)(y-128) + E(x-128)^2 + F(y-128)^2
\label{pspcorr}
\end{equation}
where $x$ and $y$ are the pixel coordinates in the BCD frame, and the
centers of the pixels run from 1 to 256 in both axes.  The convention used
by \citet{reach05} for this correction in the absolute calibration was to
define it relative to pixel (128,128).  The polynomial coefficients of the
fit are given in Table \ref{tbl-1}.  The fit was performed relative to pixel (128,128),
and then the array of correction values was normalized so 
that the median value of the correction factor over the array is 1. 
In the case of AORs designed to obtain photometry for a source using a small
dither pattern near the center of the array, the correction is near 1.0
in all channels.
A cubic fit did not improve the quality of the correction significantly.  
The fitted surfaces are shown in Figure \ref{pointcorr}, and the
coefficients are listed in Table \ref{tbl-1}.  Correction images that can
be applied to IRAC data are supplied on the SSC
website\footnote{http://ssc.spitzer.caltech.edu/irac/locationcolor/}. 
To correct the data, $F_{corr}(x,y) = F_{psp}(x,y)\times F_{measured}$.

In channels 1 and 2, the pattern is ``bowl''-shaped, with the uncorrected
photometry having the smallest value near the center of the field.  For
channels 3 and 4, the pattern is dominated by a gradient mostly left-right
across the images with opposite signs.  The maximum ranges
(maximum-minimum correction values) are 4.7\%, 5.9\%, 13\%, and 9\% for
channels 1-4, respectively.  Since the extreme values are at the edges or
corners of the arrays, the errors for uncorrected stars closer to the
centers of the arrays are much smaller. For example, for objects in the
central 128$\times$128 pixel area, the range of corrections are 1.6\%,
1.9\%, 5.6\%, and 4.0\%.

Because of the way this correction was derived, it corrects for several of
the point source gain errors at once.  These include the pixel area
difference over the FOV, the changing effective wavelength over the FOV
(see section below), and any extended/point source illumination
differences. Therefore, the correction is only strictly true for stars of
the same spectral type as the standard used in these observations.  
However, in practice the color term of the correction is relatively
smaller than the other effects, so the photometry is in general improved
when the correction is applied to all point sources.

\subsection{Effective Wavelength Variations over the Field of View}

As a consequence of the wide-field and compact optical design of the
cameras, the light at each point of the FOVs passes through the filters at
a different average angle, as shown in Figure \ref{optfig}.  In addition,
the filters are tilted with respect to the optical axis in order to
minimize abberations introduced by other optical elements and the off-axis
design.  The range in angles as a function of position results in a change
in effective wavelength over the FOV.  The filters were designed to have
the desired nominal wavelength and bandpass for the center of the field,
given the average filter tilt for the position as specified in the optical
design.  In addition to the filters, the transmission of the beamsplitters
produces angle-dependent reflection variations in channels 1 and 2, and
transmission variations in channels 3 and 4. The primary effect is a
change in the total transmission or reflection over the band as a function
of angle at the beamsplitter. The design of the filters and beamsplitters
and the measurements of their transmission and reflectance is detailed by
\citet{quijada04}.

Based on the \citet{quijada04} results, we have constructed models of the
instrument transmission for each pixel of the four channels.  The angles
of transmission through the filter and transmission or reflection of the
beamsplitter were determined for each pixel, and the total relative system
response (RSR) was calculated.  This also includes the assumed telescope
transmission and the detector quantum efficiency. Then, for a source with
a spectrum significantly different than that of the standard stars, the
color correction $K_{i,j}$ can be calculated for an object at a particular
location in an IRAC frame as described by \citet{reach05}, where the color
correction is defined as:

\begin{equation}
K_{i,j} \equiv \frac{\int{\left(F_\nu/F_{\nu_0}\right) \left(\nu/\nu_0\right)^{-1} R_{i,j} d\nu}}
              {\int{\left(\nu/\nu_0\right)^{-2} R_{i,j} d\nu}}.
\label{colcor}
\end{equation}
where $F_\nu$ is the source spectrum, $F_{\nu_0}$ is the reference spectrum
(assumed to be $\nu F_{\nu} = constant$), $R_{i,j}$ is the instrumental response
as a function of frequency at array location ({\it i,j}), and 
$\nu_0$ is the nominal frequency.

To illustrate the changes in transmission across the FOV, we have
calculated the nominal wavelength at each pixel for each channel, as
described by \citet{fazio04} for the original instrument response curves.  
The nominal wavelength was calculated using the following expression at
each pixel, integrated over the bandpass:

\begin{equation}
\lambda_0 =
\frac{\int{R d\lambda}}
       {\int{\lambda^{-1} R d\lambda}}.\label{lambda0}
\end{equation}

\begin{figure}
\epsscale{0.7}
\plotone{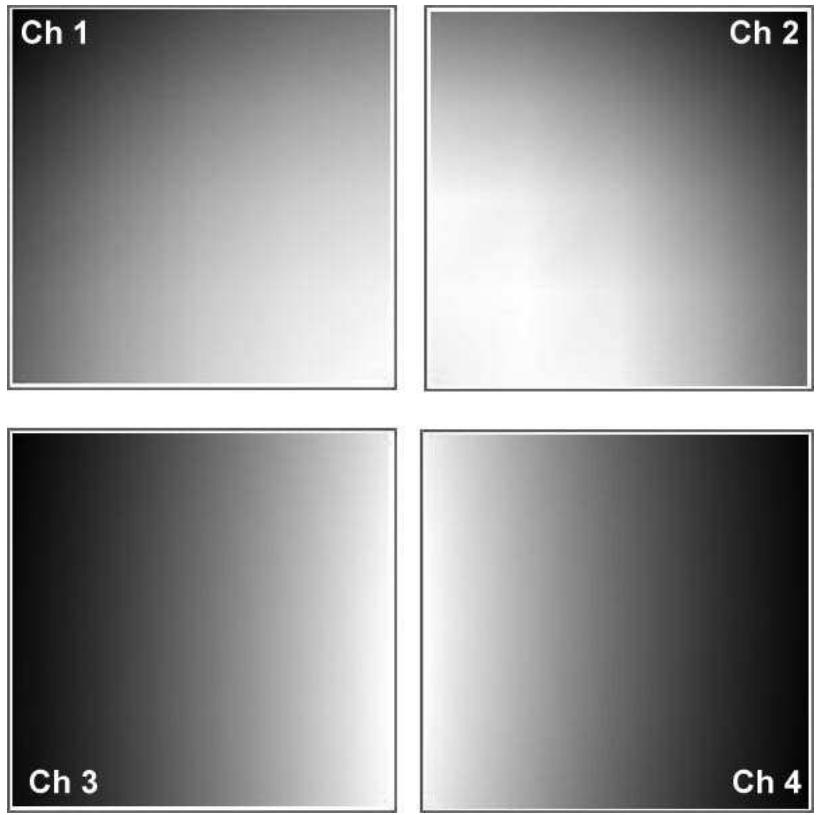}
\caption{
Changes to the nominal wavelength over the FOV for each of the IRAC
bands. The images are shown in the BCD orientation. 
Darker areas represent regions of shorter (lower) wavelength.
The nominal wavelength varies from 3.5406 to 3.5512 $\mu$m for
channel 1, 4.4680 to 4.4949 $\mu$m for channel 2, 5.6718 to 5.7458 $\mu$m
for channel 3, and 7.6212 to 7.8929 $\mu$m for channel 4.
\label{nominal}}
\end{figure}

Figure \ref{nominal} shows the variation of the nominal wavelength
across the FOV for each of the channels.  The nominal wavelength varies
from 3.5406 to 3.5512 $\mu$m for channel 1, 4.4680 to 4.4949 $\mu$m for
channel 2, 5.6718 to 5.7458 $\mu$m for channel 3, and 7.6212 to 7.8929
$\mu$m for channel 4.  The dominant change over the fields is a shift of
the entire transmission pattern, although there are some small changes in
the details of the relative response curves as one moves around the FOV,
and in Channels 3 and 4 there is a significant difference in the average
transmission over the field, mainly due to variations in the beamsplitter
transmission as noted by \citet{quijada04}.  To first order this
transmission change should be compensated for by the flat correction in
the BCD pipeline.

By applying the photometry correction described in section 2.2 above, one
is also implicitly applying a correction for the wavelength variation
effect because the correction was derived based on standard stars.  
Therefore, if a source of interest has a spectrum similar to the standard
star, no further wavelength correction is necessary.  If the source
spectrum is different from the standard, the correction for the standard
star must be backed out before the correction is applied to the data for
the specific source of interest.

\begin{figure}
\begin{center}
\includegraphics[scale=0.4,viewport=10 180 784 750,clip=true,angle=0]{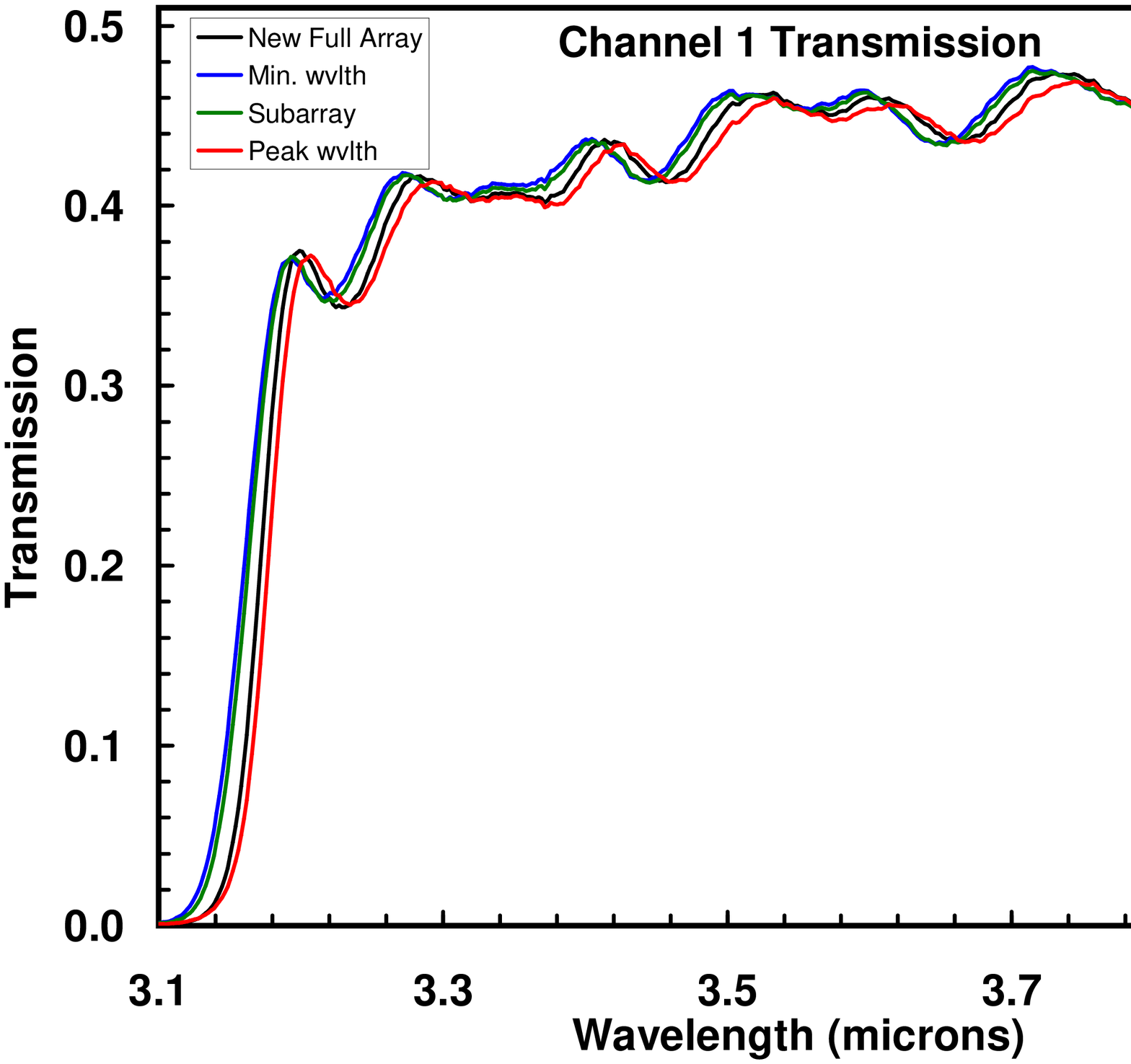}
\caption{
The instrument response function for various locations on the array for
channel 1 (``3.6 $\mu$m'').  The full array average is shown in black.  
The 32$\times$32 subarray response curve is shown in green.  The pixel
with the lowest nominal wavelength is shown in blue, and the location
with the highest wavelength in red.
 \label{RSRch1}}
\end{center}
\end{figure}

\begin{figure}
\begin{center}
\includegraphics[scale=0.4,viewport=10 180 784 750,clip=true,angle=0]{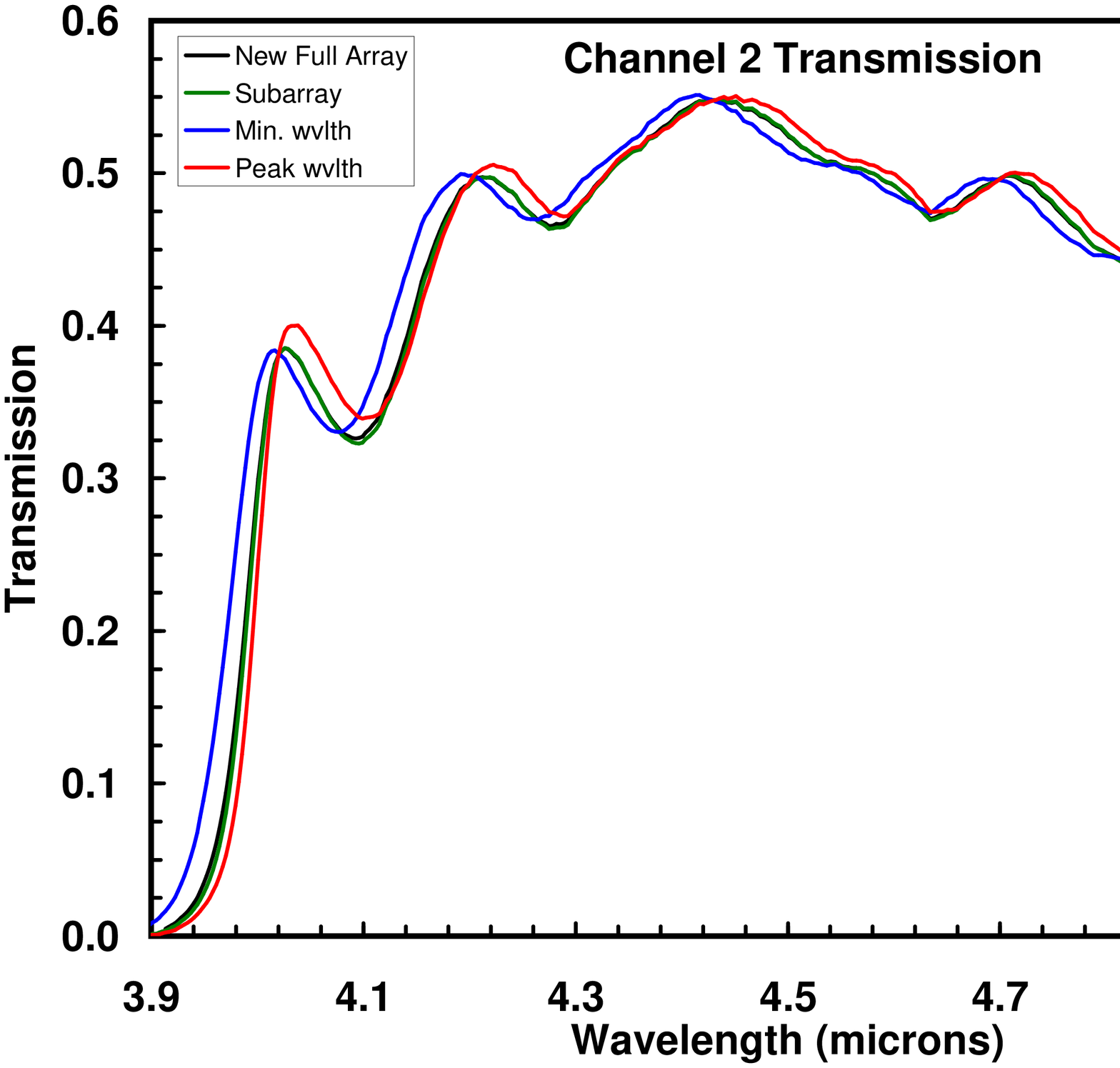}
\caption{
The instrument response function for various locations on the array for
channel 2 (``4.5 $\mu$m'').  The full array average is shown in black.  
The 32$\times$32 subarray response curve is shown in green.  The pixel
with the lowest nominal wavelength is shown in blue, and the location
with the highest wavelength in red.
 \label{RSRch2}
}
\end{center}
\end{figure}

\begin{figure}
\begin{center}
\includegraphics[scale=0.4,viewport=10 180 784 750,clip=true,angle=0]{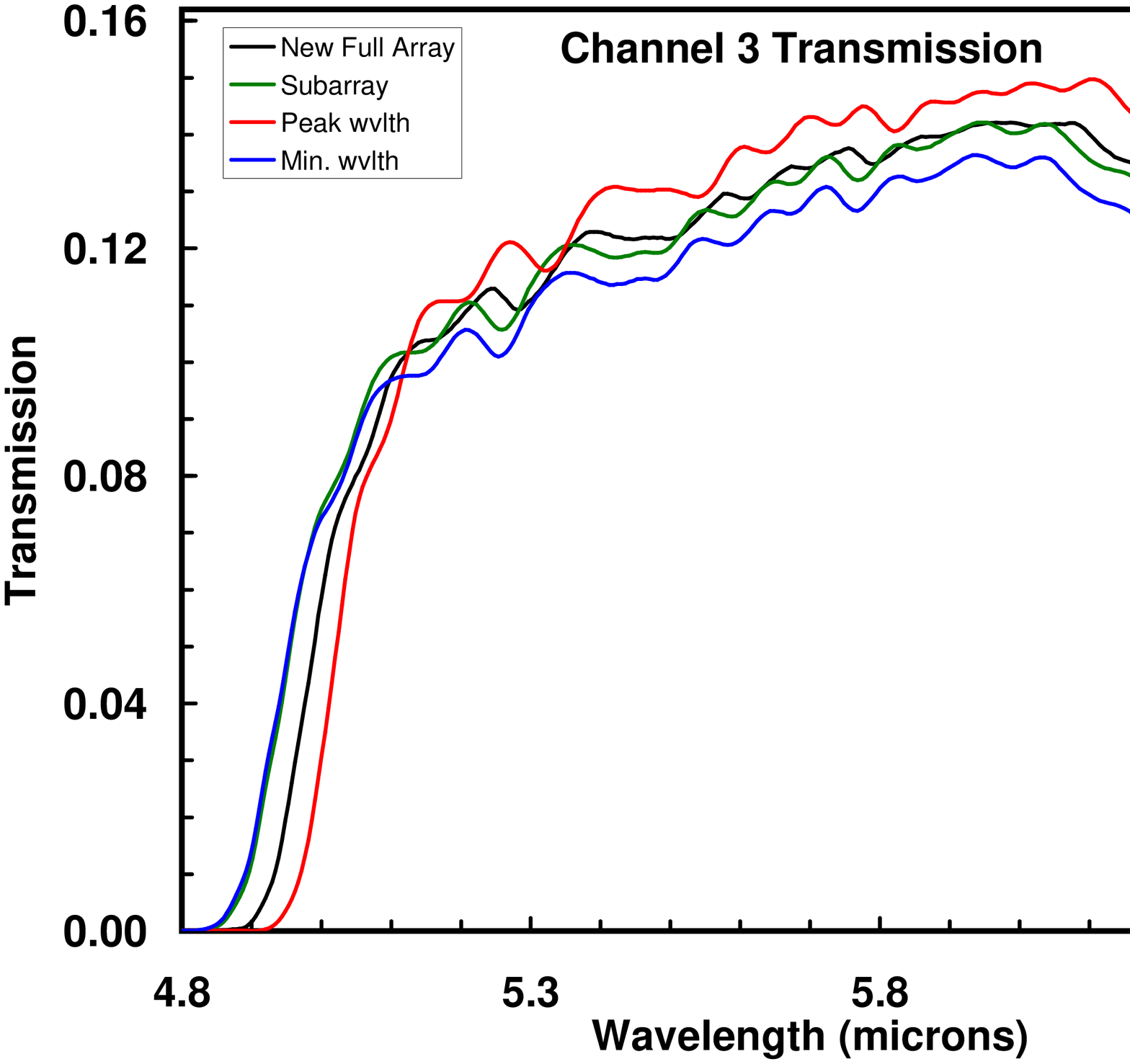}
\caption{
The instrument response function for various locations on the array for
channel 3 (``5.8 $\mu$m'').  The full array average is shown in black.  
The 32$\times$32 subarray response curve is shown in green.  The pixel
with the lowest nominal wavelength is shown in blue, and the location
with the highest wavelength in red.
 \label{RSRch3}
}
\end{center}
\end{figure}

\begin{figure}
\begin{center}
\includegraphics[scale=0.4,viewport=10 180 784 750,clip=true,angle=0]{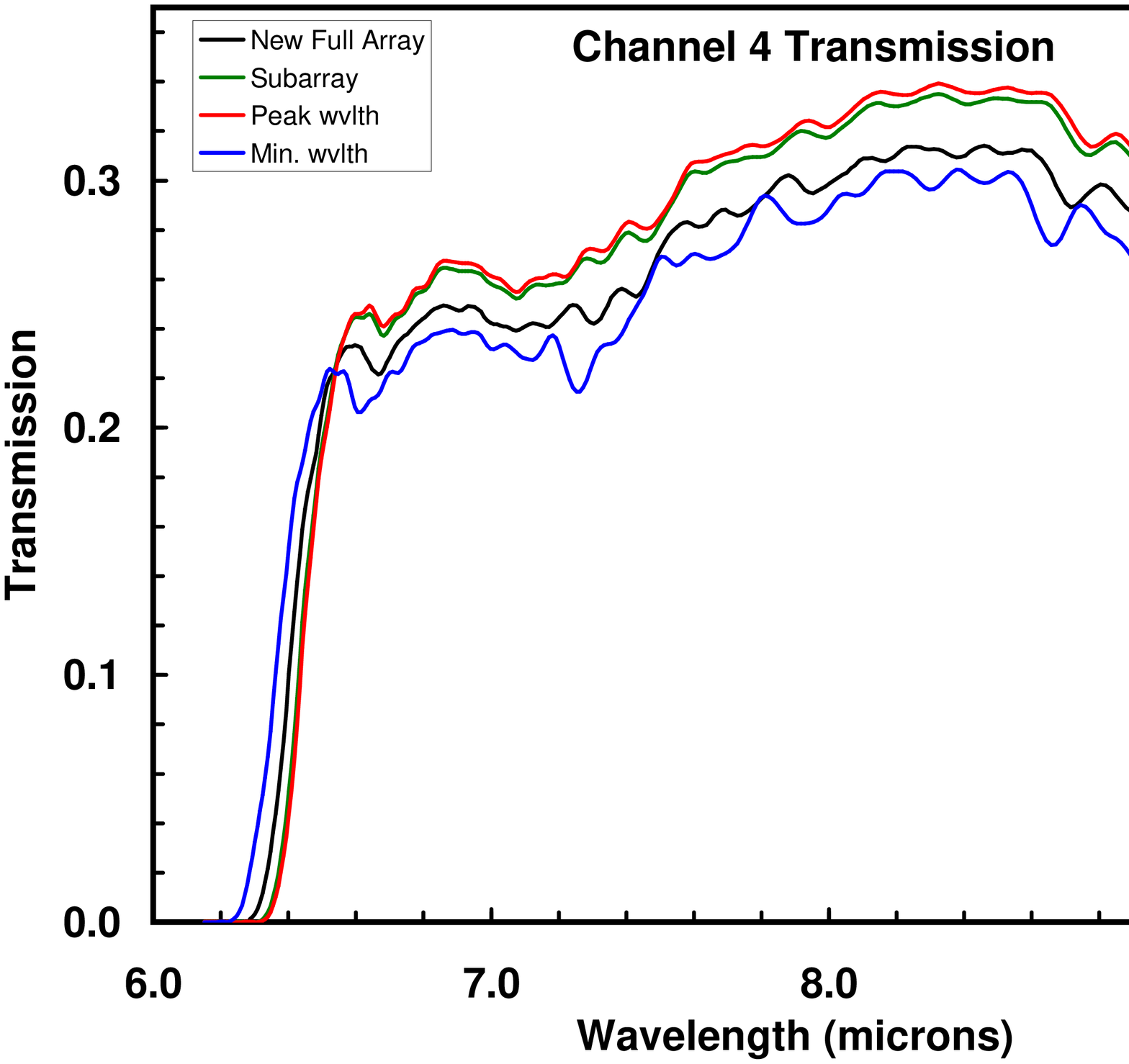}
\caption{
The instrument response function for various locations on the array for
channel 4 (``8.0 $\mu$m'').  The full array average is shown in black.  
The 32$\times$32 subarray response curve is shown in green.  The pixel
with the lowest nominal wavelength is shown in blue, and the location
with the highest wavelength in red.
 \label{RSRch4}
}
\end{center}
\end{figure}

Based on the instrument response curves for the individual pixels, we also
calculated average response curves for the entire array and for the
subarray. These are shown in Figures \ref{RSRch1} -- \ref{RSRch4}.  Each
figure shows the transmission curves averaged over the entire array, the
subarray region, and also plots of the pixels with extremes in nominal
wavelength for that channel.  These data are available on the SSC
website\footnote{http://ssc.spitzer.caltech.edu/irac}, including the full
3-d datacube with the instrument response for each pixel for all channels.

\section{Intra-pixel Gain Effects}

The optical point spread function (PSF) is slightly undersampled by the
IRAC pixel scale.  The optical model predicted image FWHM sizes of 1.6,
1.6, 1.8, and 1.9 arcsec for channels 1-4 respectively, with a pixel size
of 1.2 arcsec in all channels. The small size of the PSF causes the IRAC
photometry to be sensitive to intra-pixel gain variations, due to
variations of the quantum efficiency across a pixel area or gaps between
pixels.

\begin{figure}
\vskip 1in
\plottwo{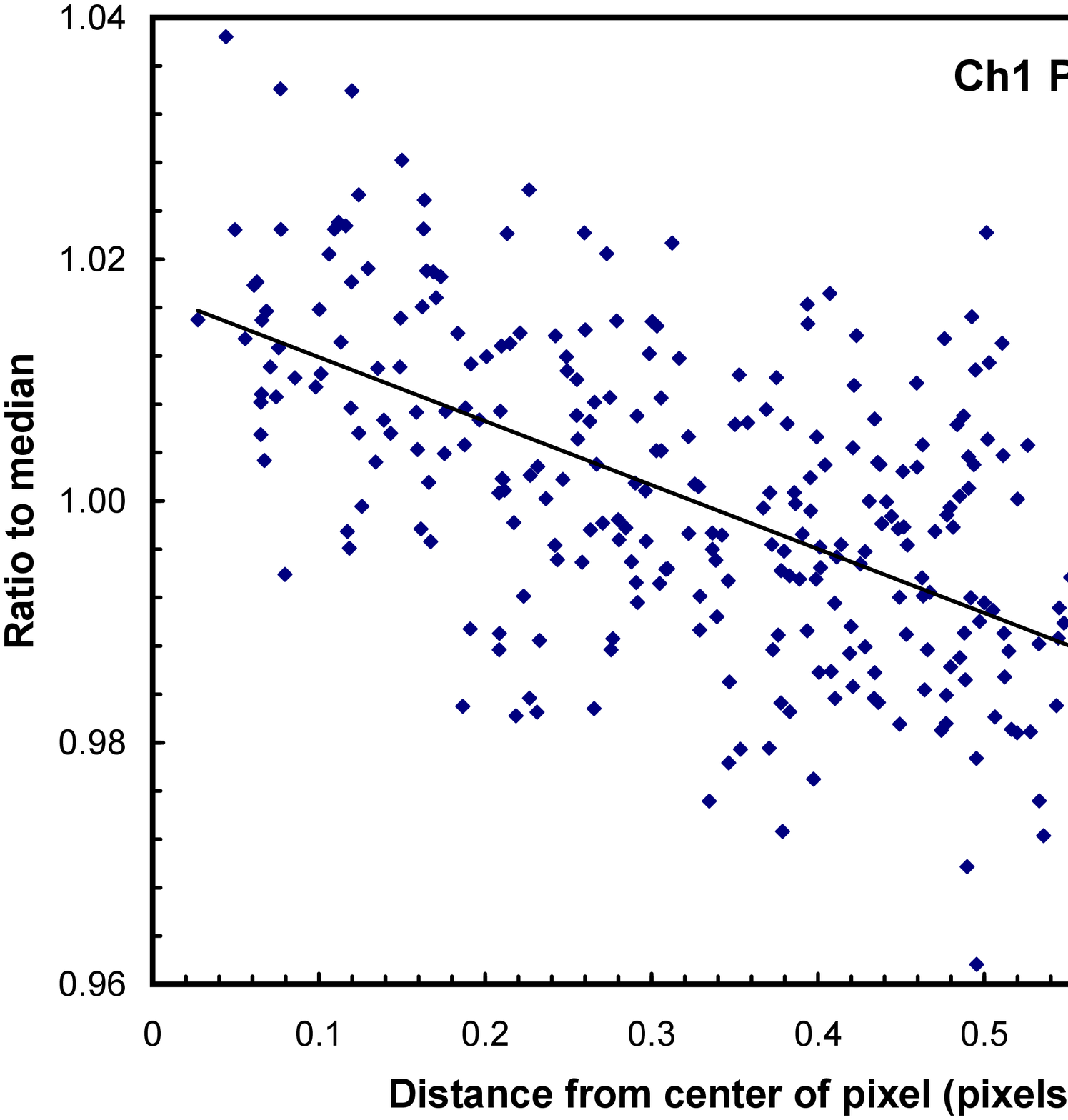}{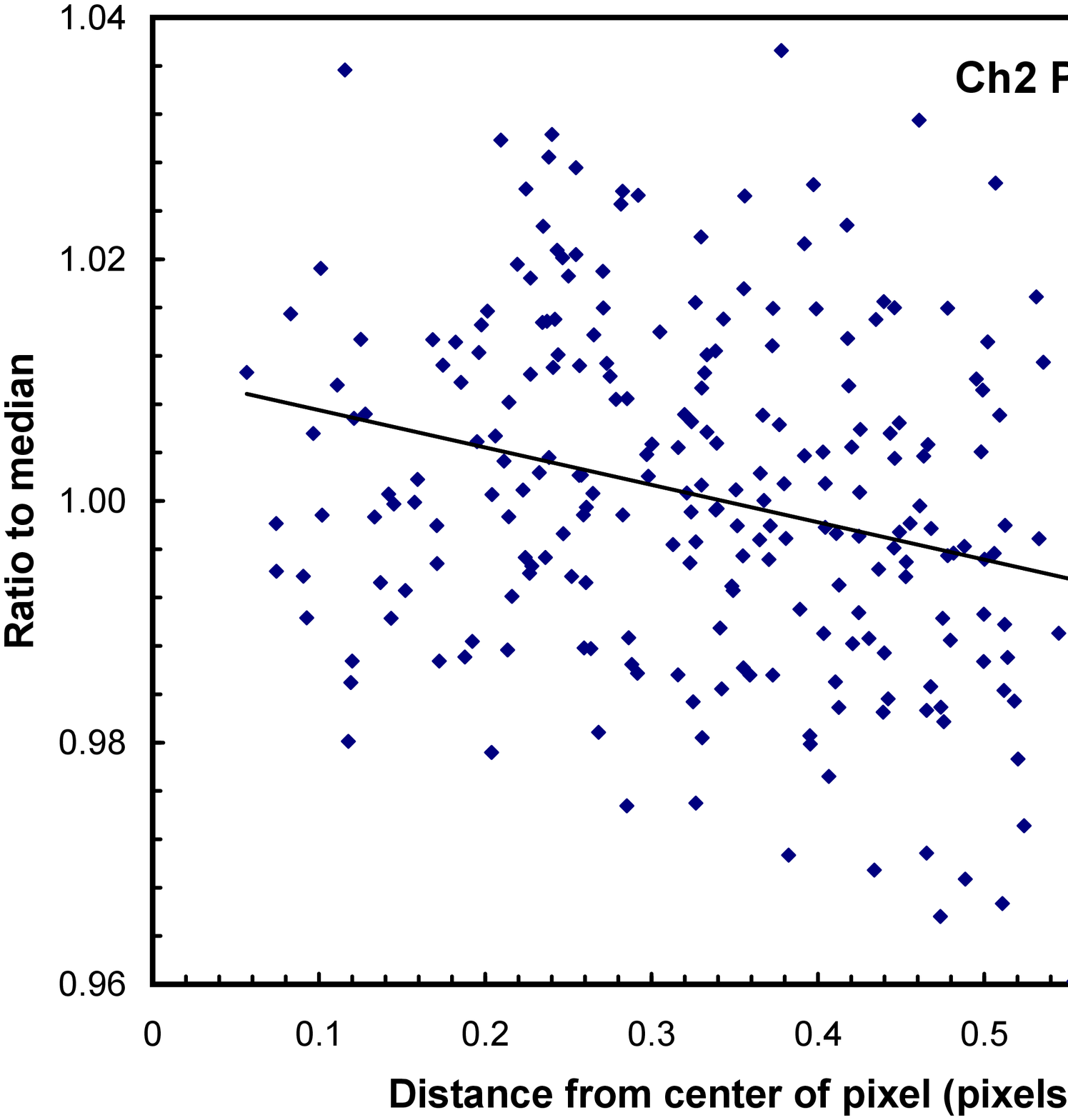}

\plottwo{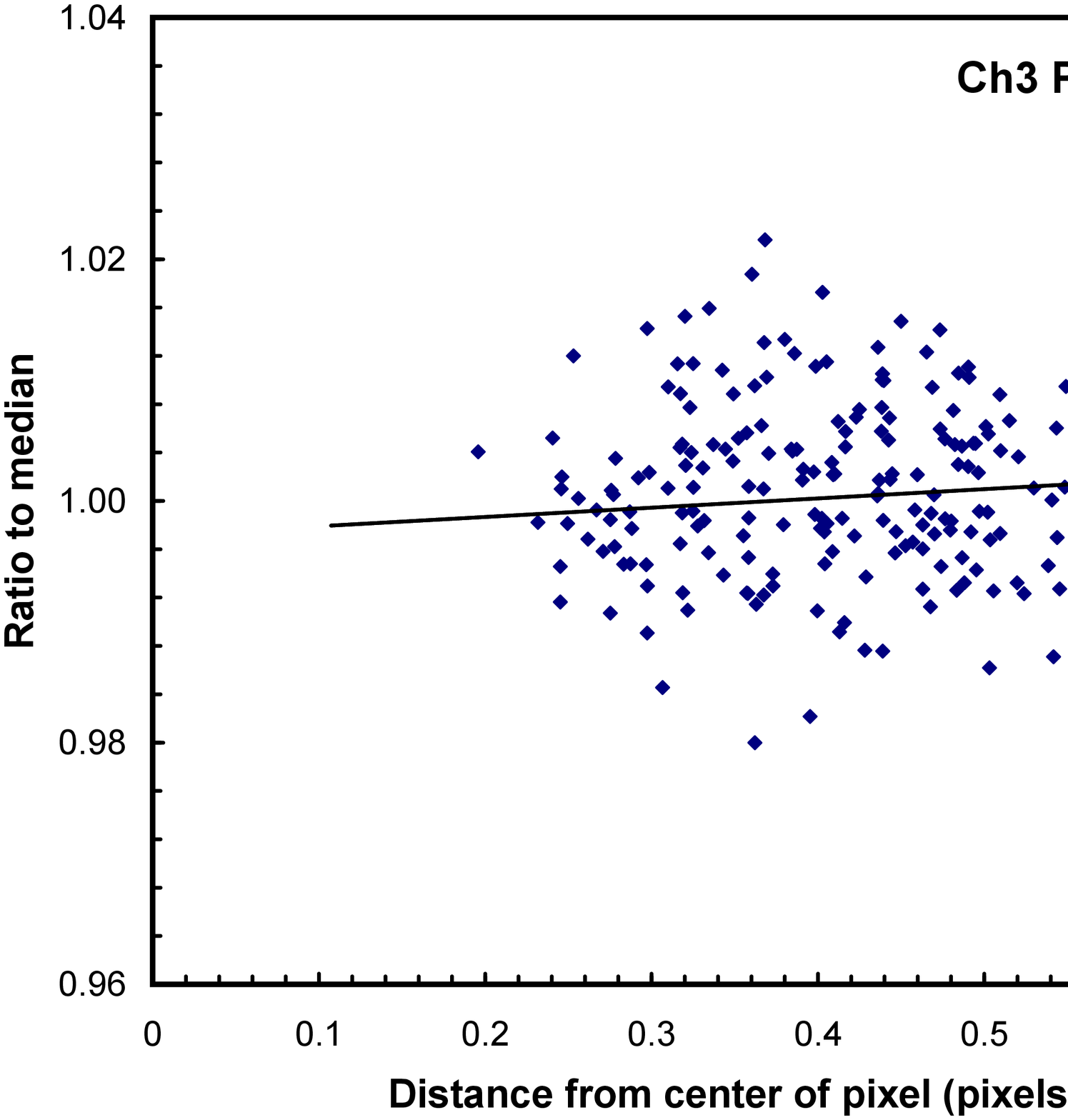}{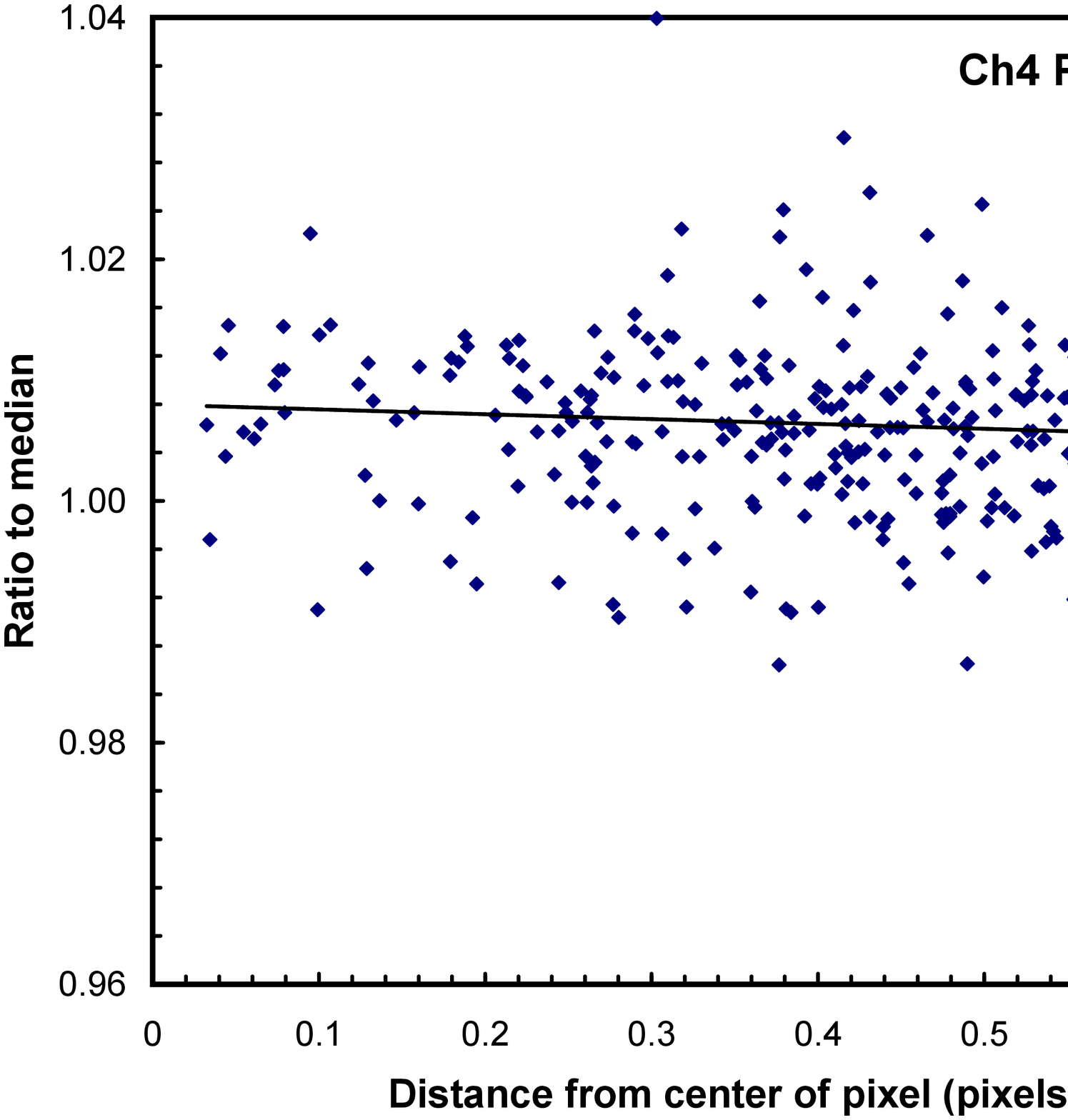}
\figcaption{
Intra-pixel gain effects on the photometry in each channel.  The extracted 
photometry of a source is plotted against the pixel phase.  The strongest 
effect is in the shortest wavelength channel, there is no significant effect
detected in channels 3 and 4 for the photometric standard star data examined.
\label{pixelphase}
}
\end{figure}

The intra-pixel gain effects were investigated by examining photometry of
stars at many different positions on the array.  The photometry was
extracted and the point source gain correction, as described in Section
2.2 above, was applied.  Then for each measurement, the photometry
relative to the median value for that star for all positions was plotted
against the distance from the source centroid to the center of the nearest
pixel.  The results are shown in Figure \ref{pixelphase}.  For channels 1
and 2, there is a correlation between the source location relative to the
pixel center (or ``pixel phase'') and the extracted photometry.  As
expected, the magnitude of this correlation is dependent on the wavelength
(or size of the PSF) -- the effect is greatest in channel 1, less in
channel 2, and not detected in channels 3 and 4.

If the correction is defined to be unity for the median location of a
source in a pixel (for randomly placed sources this location is
$1/\sqrt{2\pi}$
pixels from the center) the correction is given by 

\begin{equation}
f_{IPG}=1+A(1/\sqrt{2\pi} - p)
\end{equation}

where $p$ is the distance (in pixels) from the source centroid to the nearest
center ($0 \leq p \leq \sqrt{2}/2$).  For channel 1, $A = 0.0535$,
and for channel 2, $A = 0.0309$.

\citet{reach05} performed a correction only for channel 1, where the
effect is the largest.  The channel 1 calibration stars tended on average
to fall closer to the centers of their pixels than a random distribution,
due to the difficulty in estimating the true centroid of the flux
distribution. The median correction $\langle f_{phase}\rangle =1.0$\%.

This form of the pixel phase correction uses only one parameter, the
radial distance of the centroid from the center of the pixel.  Since the
detectors are square, a better parameterization of the effect would be
based on the x,y distance from the pixel center and also perhaps include a
model of the pixel response across the width of the pixel.  We have
derived such a correction using a simple model of the pixel response
\citep{hoffmann05}, and \citet{mighell07,mighell08} has incorporated this
into his photometry technique and demonstrated an improved correction.  
Variations in the detailed response of each individual pixel may present
the ultimate limit of how well the pixel phase can be corrected in a set
of observations.  Probably the most detailed information is available for
the pixels where the transiting planets have been observed, since this
usually involves uninterrupted periods of repeated observations of the
same source without dithering.  There is some periodic spacecraft pointing
drift that causes the source to move back and forth across neighboring
pixels.  For example, \citet{charb05} detected the pixel phase variations
in their 4.5 $\mu$m observations of the transiting planet which they were
able to reduce the residual RMS of their time series to 0.27\%.

\section{Applying the Photometry Corrections}

\subsection{Correcting photometry extracted from the BCD}

In the case where the point source is sampled at sufficient signal to
noise in a single exposure, the most straightforward approach is to apply
corrections to the BCD since the corrections are based on position in the
FOV of each frame.  For a point source in the frame, after it is extracted
using photometry software, the PSF flat field correction is applied based
on the centroid of the object.  For channels 1 and 2, the intra-pixel gain
correction can also be applied, based on the pixel phase.  If a color
correction is necessary, then the correction is calculated based on the
source spectrum and the PSF position and applied to the photometry.

\begin{figure}
\begin{center}
\includegraphics[scale=0.4,angle=-90]{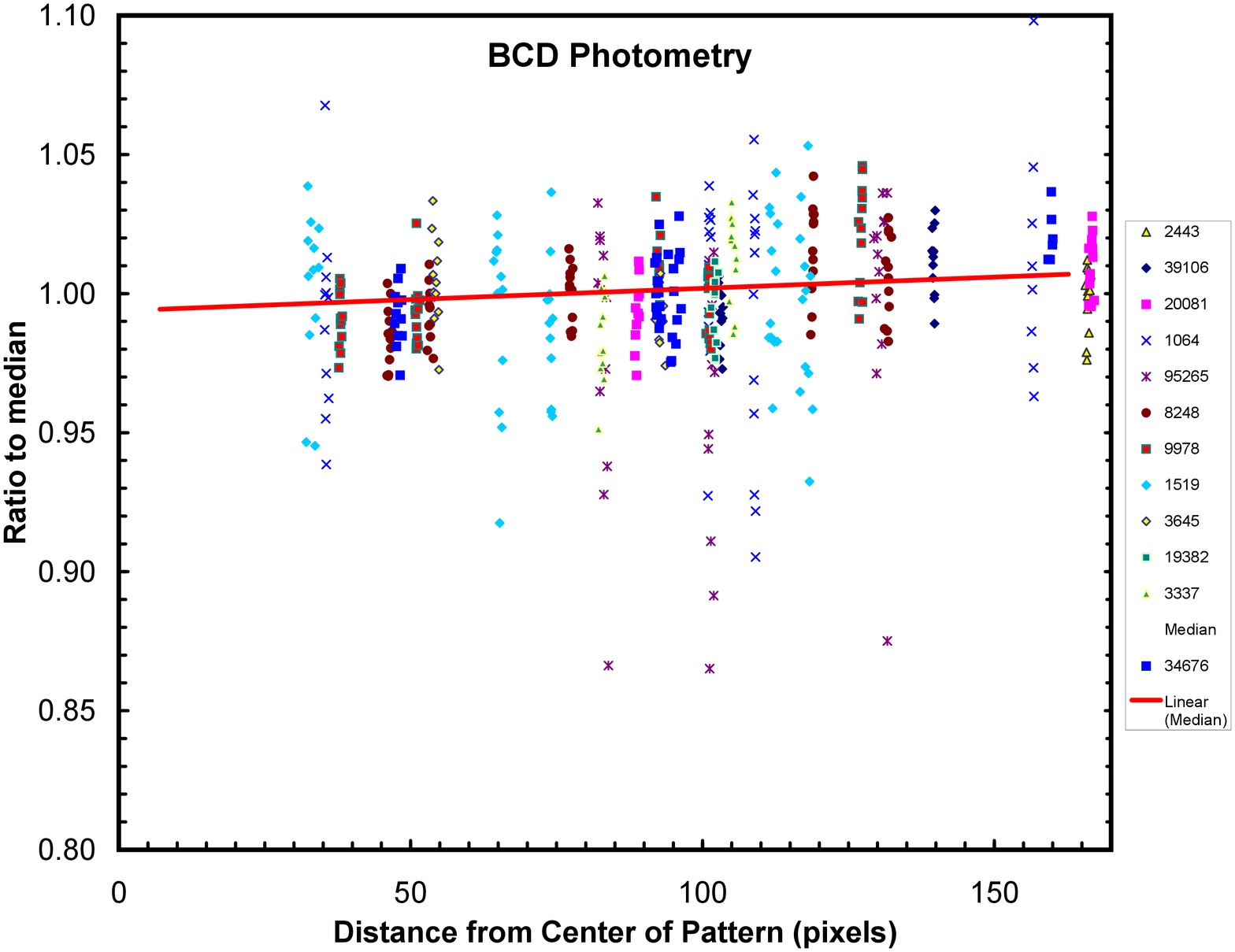}
\includegraphics[scale=0.4,angle=-90]{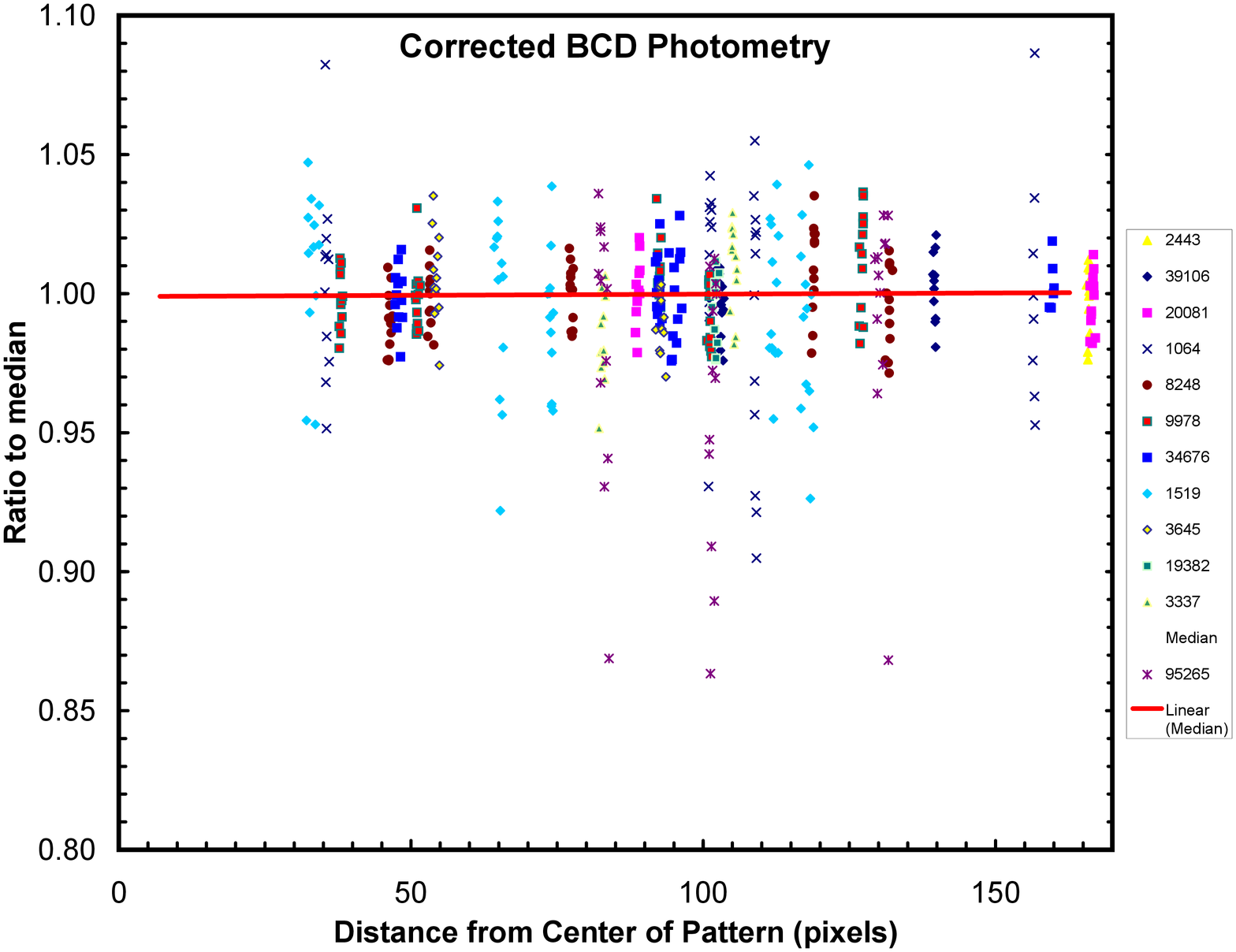}
\end{center}
\caption{
The top plot shows the relative photometry of stars in a test field 
in Channel 1
(the ratio of the measurement to the median value at that array position)
as a function of the distance from the center of the photometric 
correction pattern in Figure \ref{pointcorr}.  Individual stars have unique colors and 
symbols, several stars have multiple measurements at different array positions. 
A linear fit to the median values shows the trend of lower values towards
the center.  The lower plot shows the results after the gain corrections have been
applied.  The overall trend with array position is removed, and the scatter
of individual positions is reduced.\label{photcorr}
}
\end{figure}

\begin{figure}
\begin{center}
\includegraphics[scale=0.4,viewport=10 180 784 750,clip=true,angle=0]{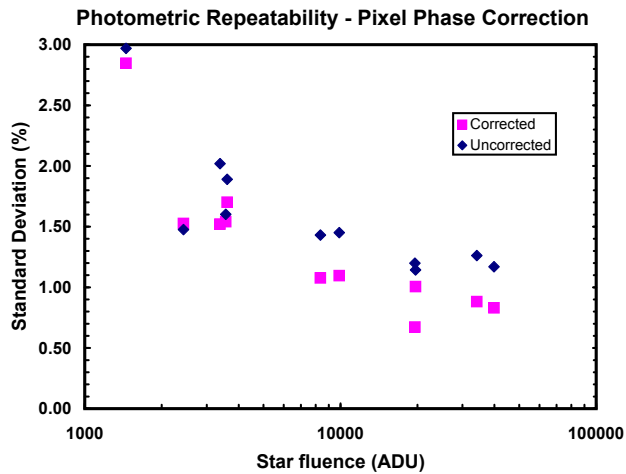}
\end{center}
\caption{
The standard deviation of photometry of several stars observed in Channel
1, plotted against the fluence in ADUs. The dark blue points are
uncorrected, the magenta points are after the intrapixel gain correction
has been applied.  The standard deviations are reduced by about 0.5\% for
most stars. \label{noise}}
\end{figure}

Figure \ref{photcorr} shows an example of a test case where the pixel
phase and point source gain corrections were applied. The top plot shows
the relative Channel 1 photometry of stars as a function of distance from
the center of the photometric correction pattern in Figure
\ref{pointcorr}.  A linear fit to the data illustrates the trend of lower
photometry near the center of the pattern. The second plot shows the same
data after the gain corrections have been applied.  The linear fit to the
relative photometry now shows no trend with position on the array, and the
scatter for many of the positions is noticeably lower.  Figure \ref{noise}
shows the effect of the phase correction on the photometry.  For stars of
several fluences, the standard deviation of the photometry at a particular
point on the array was calculated from the available measurements.  Also
shown are the results from the same data after the intrapixel gain
correction was applied. The reduction in the scatter of the photometry was
about 0.5\% for the stars examined.

\subsection{Correcting photometry extracted from mosaics}

Correcting photometry extracted from mosaics is more complicated because
the image at any point source location is the combination of images at
several different FOV positions in the individual frames.  The pixels
contain some mixture of extended and point source emission. Therefore if
one applies the correction before mosaicing, the correction is wrongly
applied to the extended emission, and that will create artifacts in the
images.

One possible way to apply the point source gain correction to mosaics is
to extract the photometry from the uncorrected mosaic and also to
calculate a separate correction map.  The correction map is calculated by
successively offsetting the BCD correction image to the same locations as
the science images and calculating the mean corrections at each location
in the final mosaic. The BCD correction image, however, includes the
correction for the change in pixel area over the array, and the mosaics
have been reprojected to a constant pixel area.  Therefore we have to
divide out the pixel area normalization from the BCD correction image
before we use it to make the mosaic correction map.

\begin{figure}
\epsscale{1}
\plotone{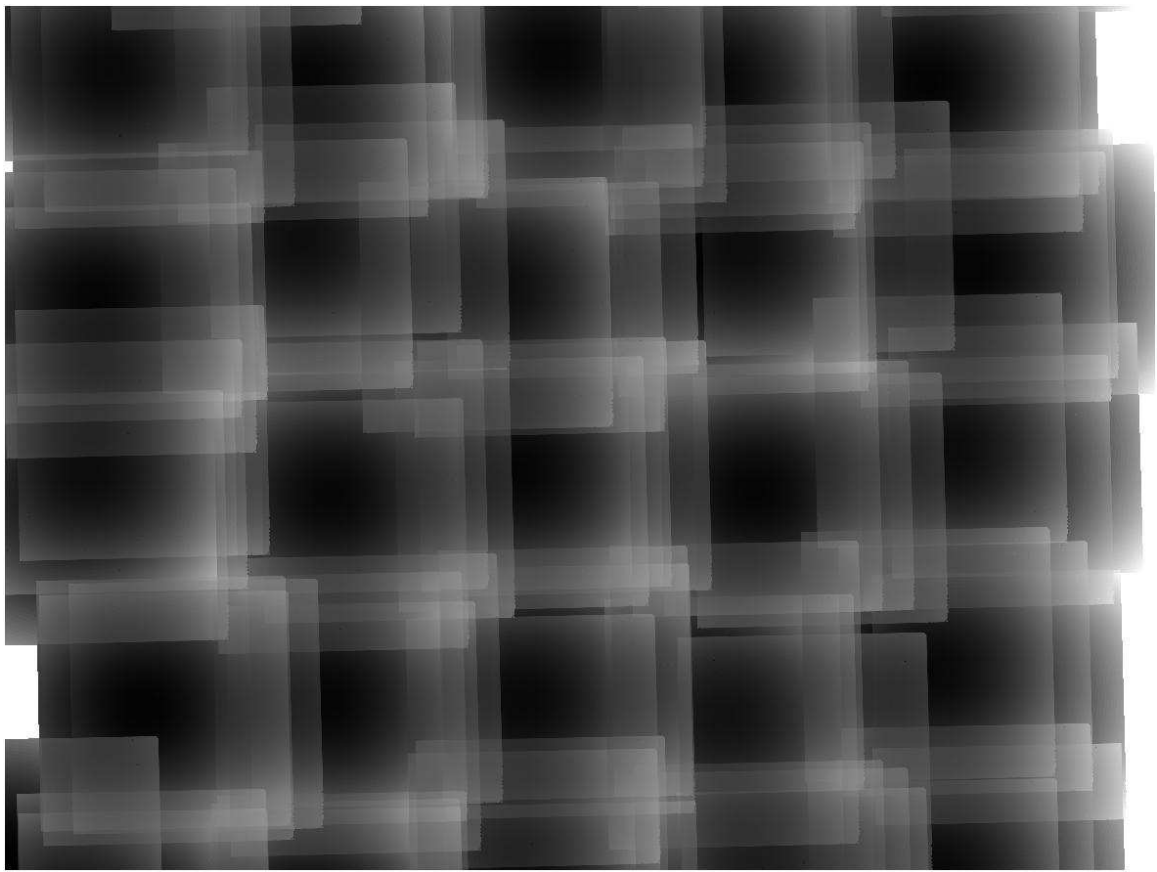}
\caption{Point source photometry correction image for a channel 1 mosaic
made from a mapping AOR that used three dithers per map position.  The
range of correction values is -1.2\% (the dark regions) to +3.4\% (the
ligher areas).  \label{mosaiccorr} }
\end{figure}

An example of such a correction map for a mapping AOR with dithers is
shown in Figure \ref{mosaiccorr}.  For each source, one would need to
determine its pixel coordinates and use the value at the same location in
the correction mosaic to correct the photometry. This approach will work
in cases where either the target is of high surface brightness relative to
the local extended emission, including the zodi, or the extended emission
has a spectrum similar to the zodi emission.  The wavelength-dependent
part of the correction depends on the spectral slope of the sum of the
extended and pointlike emission (i.e. what actually arrived at the
detector), not the spectral slope of the point source alone. If the local
extended-source spectrum differs from the zodi, the extended emission is
also in need of correction, and so the method will have problems in
proportion to the strength of the extended emission.

The correction mosaics themselves will have seams all through them as seen
in Figure \ref{mosaiccorr}, and it isn't obvious for an object whose
photometry is derived from many pixels just how to derive a single
correction factor applicable to the catalog entry except by duplicating
the extraction procedure. The correction would be even more difficult to
derive for PSF fitting. A PSF-weighted average of the pixels in the
correction mosaic might closely approximate the correction value for a
particular position in the mosaic.

For a data set such as an extragalactic survey like SWIRE that has a
relatively uniform background and few bright, extended emission sources,
the data can be split into a nearly constant ``background'' and residual
``object'' components, and the corrections in image space applied to the
``object image''. Then the zodi-flattened (zodi-colored)  background can
be put back in. The resulting mosaics would be seamless and fully
calibrated for either point sources or extended emission. However, this
technique is possible only where the background is uniform and has a color
similar to the zodi emission. A fully generic implementation, which is
under development at the SSC, will involve pixel-wise corrections that
take into account the pixel colors and thus be correct for both point
sources and extended emission.

\subsection{Aperture corrections}

The IRAC calibration described by \citet{reach05} used a 10 pixel aperture
and an aperture with an inner radius of 12 pixels and outer radius of 20
pixels. This is appropriate for measuring standard stars since they were
chosen to be extremely bright relative to background objects within this
distance on the sky.  The aperture contains a large fraction of the total
flux from the object, but is small enough so that it does not extend off
the array when extracting the photometry from BCDs where the star is
offset at different positions on the array.  However, for many other
applications where one is extracting sources in crowded fields or where
there is significant extended emission near the source, a smaller aperture
is likely to produce more accurate photometry.  In order to calibrate the
photometry using aperture sizes different from that of the standard star
observations, one must perform an extraction of the same star(s) using the
different parameters, and a correction factor can then be determined.  
This analysis was performed based on observations of a standard star at
many different positions on the array. The procedure {\tt aper} in IDLPHOT
was used to extract the photometry from the BCD, and the average
corrections were determined.  The results are reported in the IRAC Data
Handbook\footnote{http://ssc.spitzer.caltech.edu/irac/dh/} (IDH; see Table
5.7 of the Handbook). Note that for extracting photometry from the BCD,
there might be a position dependence due to the distortion, especially for
small apertures.  This has not been taken into account in the IDH where
the average value for all positions with the same pixel aperture
dimensions are reported.

The aperture correction values in the IDH were determined from the BCD of
a star with high S/N observations and many positions on the array.  Many
observers instead extract photometry from mosaics for which the number of
dithers at each position and the reprojected pixel size is different from
the instrumental pixels. The data set and the choices one makes in the
reduction parameters can make a difference in the aperture photometry
corrections.  In order to illustrate this, we measured the aperture
corrections for two different datasets using a few different mosaic
parameters.  The datasets that were used were an AOR from the
Extragalactic First Look Survey \citep[FLS;][-
\dataset{ADS/Sa.Spitzer\#0003863296}]{lacy05}, and a mosaic made from the
SAGE data \citep{meixner06} contained in an area of 0.7x0.7 deg, centered
on 77.5d R. A., -65.166667 Dec. (J2000.0).  The FLS AOR used a mapping
pattern with 5x100s dithers at each map point.  The SAGE data were taken
with a coverage of 2x12s HDR frames in two separate epochs separated by
approximately 3 months, so at each position the depth of coverage is at
least 4 frames with two different rotation angles roughly 90$^{\circ}$
apart.  The location of the SAGE mosaic was chosen to be in a region of
good coverage in a corner of the field, far from the dense LMC stellar
distribution and any extended emission.

The data for each survey field were mosaicked to a pixel scale of
0\farcs6/pixel using IRACproc \citep{schuster06}, which is based on the
mopex software produced by the SSC \citep{makovoz06}. We also used the
post-BCD product generated by the SSC for the FLS data, which is at a
scale of 1\farcs2/pixel.  For each mosaic, we used the {\tt daofind} and
{\tt phot} tasks in IRAF to find and extract photometry for the sources in
the field.  We used only those sources with a S/N$>$100, and also used a
minimum flux cutoff to exclude the fainter sources.  The same sources were
extracted using the same range of apertures and background annulus sizes
that were used to produce the table of corrections in the IDH.  The ratio
of fluxes relative to the photometry using a 10 pixel aperture with a
background annulus range from 10-20 pixels were calculated, and the median
ratio was determined for each set of parameters.  The estimated
uncertainties of the ratios are approximately 0.002, 0.003, 0.007, and
0.008 in channels 1 - 4, respectively.  The results are summarized for
each channel in Tables \ref{ch1appcor} - \ref{ch4appcor}.

The first column of each table shows the IDH values for comparison.  The
values are very similar for channels 3 and 4; the largest variations are
in channel 1 and 2.  Several effects could influence the correction
factors, including the pixel scale of the mosaics (as shown in the FLS 0.6
and PBCD results), differences in mosaic mapping and dithering techniques
and depths (compare the LMC 0.6 to the FLS 0.6), and software that was
used to extract the photometry (compare the IDH values to all others
reported here).  For the most accurate aperture correction factors, one
should determine the corrections based on the data that one is performing
the photometry on, or on a data set taken with a similar observing
strategy.

\section{Conclusions}

The IRAC camera has FOV-dependent transmission characteristics that affect
the measurement of astronomical sources.  After correction of these
effects for standard stars, \citet{reach05} found that the calibration has 
a relative 
accuracy of 1.8\%, 1.9\%, 2.0\%, and 2.1\% in channels 1 (3.6 $\mu$m), 2
(4.5 $\mu$m), 3 (5.8 $\mu$m), and 4 (8 $\mu$m), respectively.  To measure
fluxes at this level of accuracy requires several photometric corrections:
array position dependence (due to changing spectral response and pixel
solid angle over the camera of view), pixel phase dependence (due to
nonuniform quantum efficiency over a pixel), color correction (due to the
different system response integrated over the passband for sources of
different color), and aperture correction (due to the fractions of light
included within the measurement aperture and lost in the background
aperture).  The same accuracies are possible for sources with spectra
similar to the A-type standards used in the absolute calibration with the
array position dependence, pixel phase, and aperture corrections.  For
sources with spectra different from A stars, a knowledge of the source
spectrum is necessary to make the necessary color corrections to achieve
the same photometric accuracy.

\acknowledgements

This work is based on observations made with the Spitzer Space Telescope,
which is operated by the Jet Propulsion Laboratory, California Institute
of Technology under NASA contract 1407. Support for this work was provided
by NASA through an award issued by JPL/Caltech.

{\it Facilities:} \facility{Spitzer (IRAC)}

\clearpage

\clearpage

\begin{table}
\begin{center}
\caption{IRAC Photometry correction fit coefficients.\label{tbl-1}}
\begin{tabular}{crrrrrr}
\tableline\tableline
IRAC & & &Coefficients\tablenotemark{a}\\
Channel & $A$ & $B$ & $C$ & $D$ & $E$ & $F$ \\
\tableline
1 & 1.0114 & -3.536E-6 & -6.826E-5 & -1.618E-8 & 1.215E-6 & 1.049E-6\\
2 & 1.0138 & 8.401E-5 & 3.345E-7 & 1.885E-7 & 1.438E-6 & 1.337E-6\\
3 & 1.0055 & -3.870E-4 & 4.600E-5 & 1.956E-7 & 2.078E-6 & 9.970E-7\\
4 & 1.0054 & 2.332E-4 & -8.234E-5 & -1.881E-7 & 6.520E-7 & 9.415E-7\\
\tableline
\end{tabular}
%% Any table notes must follow the \end{tabular} command.
\tablenotetext{a}{The coefficient labels are defined in  equation \ref{pspcorr} 
}
%\tablecomments{We can also attach a long-ish paragraph of explanatory
\end{center}
\end{table}

\begin{table}
\begin{center}
\caption{IRAC Channel Transmission Summary\label{tbl-2}}
\begin{tabular}{ccccccc}
\tableline\tableline
IRAC & Nominal& Central\tablenotemark{a}& & & Average\tablenotemark{c}  \\
Channel & Wavelength & Wavelength & Bandpass & Bandpass & Transmission \\
 & ($\mu$m) & ($\mu$m) & ($\mu$m) & (percent) & \\
\tableline
1 & 3.544 & 3.543 & 0.747 & 21.1 & 0.430 \\
2 & 4.479 & 4.501 & 1.018 & 22.7 & 0.469 \\
3 & 5.710 & 5.711 & 1.412 & 24.8 & 0.125 \\
4 & 7.844 & 7.905 & 2.910 & 37.0 & 0.280 \\
Subarray:\\
1 & 3.534 & 3.538 & 0.740 & 20.9 & 0.426 \\
2 & 4.489 & 4.506 & 1.009 & 22.5 & 0.457 \\
3 & 5.679 & 5.687 & 1.383 & 24.3 & 0.122 \\
4 & 7.884 & 7.912 & 2.861 & 36.4 & 0.289 \\
\tableline
\end{tabular}
%% Any table notes must follow the \end{tabular} command.
\tablenotetext{a}{The central wavelength is defined as the midpoint between
the half-power points of the bandpass.  The half-power points are defined as 
being the wavelengths where the transmission is 50% of the peak filter 
transmission.}
\tablenotetext{b}{The bandpass is the distance in wavelength
between the half-power points of the transmission.}
\tablenotetext{c}{The average transmission is determined by averaging the 
transmission between the half-power points of the bandpass.}
\end{center}
\end{table}

\begin{table}
\begin{center}
\caption{Channel 1 Aperture Correction Factors\label{ch1appcor}}
\begin{tabular}{ccrrrr}
\tableline\tableline
aperture & annulus\\
radius\tablenotemark{a} & range &IDH\tablenotemark{b}&	FLS 0.6\tablenotemark{c}& 
FLS PBCD\tablenotemark{d}& 	LMC 0.6\tablenotemark{e} \\
\tableline
5 & 10-20 & 1.049	&1.049&	1.057	&1.057\\
5 & 5-10 & 1.061	&1.057&	1.066	&1.068\\
3 & 10-20 & 1.112	&1.101&	1.124	&1.128\\
3 & 3-7 & 1.124	&1.111&	1.138	&1.141\\
2 & 10-20 & 1.205	&1.167&	1.254	&1.243\\
2 & 2-6 & 1.213	&1.174&	1.263	&1.251\\
\tableline
\end{tabular}
\tablenotetext{a}{aperture and annulus ranges are in native pixel units (1\farcs22).}
\tablenotetext{b}{IRAC Data Handbook values}
\tablenotetext{c}{First Look Survey data, 0.6 arcsec/pixel mosaic}
\tablenotetext{d}{First Look Survey data, 1.2 arcsec/pixel post-BCD mosaic}
\tablenotetext{e}{SAGE LMC data, 0.6 arcsec/pixel mosaic, both epochs}
\end{center}
\end{table}

\begin{table}
\begin{center}
\caption{Channel 2 Aperture Correction Factors\label{ch2appcor}}
\begin{tabular}{ccrrrr}
\tableline\tableline
aperture & annulus\\
radius\tablenotemark{a} & range &IDH\tablenotemark{b}&	FLS 0.6\tablenotemark{c}& 
FLS PBCD\tablenotemark{d}& 	LMC 0.6\tablenotemark{e} \\
\tableline
5 & 10-20 & 1.050&1.059&1.059&1.063\\
5 & 5-10 & 1.064&1.074&1.074&1.077\\
3 & 10-20 & 1.113&1.124&1.126&1.129\\
3 & 3-7 & 1.127&1.136&1.139&1.141\\
2 & 10-20 & 1.221&1.250&1.266&1.262\\
2 & 2-6 & 1.234&1.263&1.282&1.277\\
\tableline
\end{tabular}
\tablenotetext{a}{aperture and annulus ranges are in native pixel units (1\farcs22).}
\tablenotetext{b}{IRAC Data Handbook values}
\tablenotetext{c}{First Look Survey data, 0.6 arcsec/pixel mosaic}
\tablenotetext{d}{First Look Survey data, 1.2 arcsec/pixel post-BCD mosaic}
\tablenotetext{e}{SAGE LMC data, 0.6 arcsec/pixel mosaic, both epochs}
\end{center}
\end{table}

\begin{table}
\begin{center}
\caption{Channel 3 Aperture Correction Factors\label{ch3appcor}}
\begin{tabular}{ccrrrr}
\tableline\tableline
aperture & annulus\\
radius\tablenotemark{a} & range &IDH\tablenotemark{b}&	FLS 0.6\tablenotemark{c}& 
FLS PBCD\tablenotemark{d}& 	LMC 0.6\tablenotemark{e} \\
\tableline
5 & 10-20 & 1.058&1.055&1.056&1.055\\
5 & 5-10 & 1.067&1.066&1.068&1.064\\
3 & 10-20 & 1.125&1.127&1.135&1.129\\
3 & 3-7 & 1.143&1.148&1.155&1.147\\
2 & 10-20 & 1.363&1.370&1.391&1.386\\
2 & 2-6 & 1.379&1.385&1.408&1.402\\
\tableline
\end{tabular}
\tablenotetext{a}{aperture and annulus ranges are in native pixel units (1\farcs22).}
\tablenotetext{b}{IRAC Data Handbook values}
\tablenotetext{c}{First Look Survey data, 0.6 arcsec/pixel mosaic}
\tablenotetext{d}{First Look Survey data, 1.2 arcsec/pixel post-BCD mosaic}
\tablenotetext{e}{SAGE LMC data, 0.6 arcsec/pixel mosaic, both epochs}
\end{center}
\end{table}

\begin{table}
\begin{center}
\caption{Channel 4 Aperture Correction Factors\label{ch4appcor}}
\begin{tabular}{ccrrrr}
\tableline\tableline
aperture & annulus\\
radius\tablenotemark{a} & range &IDH\tablenotemark{b}&	FLS 0.6\tablenotemark{c}& 
FLS PBCD\tablenotemark{d}& 	LMC 0.6\tablenotemark{e} \\
\tableline
5 & 10-20 & 1.068	&1.063&1.065&1.065\\
5 & 5-10 & 1.089&1.082&1.085&1.087\\
3 & 10-20 & 1.218&1.217&1.233&1.234\\
3 & 3-7 & 1.234&1.233&1.249&1.248\\
2 & 10-20 & 1.571&1.569&1.587&1.597\\
2 & 2-6 & 1.584&1.585&1.602&1.609\\
\tableline
\end{tabular}
\tablenotetext{a}{aperture and annulus ranges are in native pixel units (1\farcs22).}
\tablenotetext{b}{IRAC Data Handbook values}
\tablenotetext{c}{First Look Survey data, 0.6 arcsec/pixel mosaic}
\tablenotetext{d}{First Look Survey data, 1.2 arcsec/pixel post-BCD mosaic}
\tablenotetext{e}{SAGE LMC data, 0.6 arcsec/pixel mosaic, both epochs}
\end{center}
\end{table}

\end{document}